# Abundance Enhancements in Impulsive Solar Energetic-Particle Events with Associated Coronal Mass Ejections

**Donald V. Reames[1], Edward W. Cliver[2], and Stephen W. Kahler[3]**

[1]Institute for Physical Science and Technology, University of Maryland, College Park, MD 20742-2431 USA, email: dvreames@umd.edu

[2]Space Vehicles Directorate, Air Force Research Laboratory, Sunspot, NM 88349, USA, email: ecliver@nso.edu

[3]Air Force Research Laboratory, Space Vehicles Directorate, 3550 Aberdeen Avenue, Kirtland AFB, NM 87117, USA, email: stephen.kahler@kirtland.af.mil

**Abstract** We study the abundances of the elements He through Pb in Fe-rich impulsive solar energetic-particle (SEP) events with measurable abundances of ions with atomic number $Z>2$ observed on the *Wind* spacecraft, and their relationship with coronal mass ejections (CMEs) observed by the *Large Angle and Spectrometric Coronagraph* (LASCO) onboard the *Solar and Heliospheric Observatory* (SOHO). On average the element abundances in these events are similar to coronal abundances at low $Z$ but, for heavier elements, enhancements rise as a power law in the mass-to-charge ratio $A/Q$ of the ions (at coronal temperatures of 2.5–3 MK) to a factor of 3 at Ne, 9 at Fe, and 900 for $76 \leq Z \leq 82$. Energy dependences of abundances are minimal in the 2–15 MeV amu$^{-1}$ range. The 111 of these Fe-rich impulsive SEP events we found, between November 1994 and August 2013 using the *Wind* spacecraft, have a 69% association rate with CMEs. The CMEs are narrow with a median width of 75$^{\circ}$, are characteristically from western longitudes on the Sun, and have a median speed of ≈600 km s$^{-1}$. Nearly all SEP onsets occur within 1.5–5 h of the CME onset. The faster (>700 km s$^{-1}$), wider CMEs in our sample are related to SEPs with coronal abundances indicating hot coronal plasma with fully ionized He, C, N and O and moderate enhancements of heavier elements, relative to He, but slower (<700 km s$^{-1}$), narrower CMEs emerge from cooler plasma where higher SEP mass-to-charge ratios, $A/Q$, yield much greater abundance enhancements, even for C/He and O/He. Apparently, the open magnetic-reconnection region where the impulsive SEPs are accelerated also provides the energy to drive out CME plasma, accounting for a strong, probably universal, impulsive SEP-CME association.

# 1. Introduction

The relative abundances of the chemical elements has been one of our most powerful tools in determining the identity and the physical processes of acceleration of a variety of energetic ion populations seen throughout the heliosphere (*e.g.* Reames 1999). Ever since the pioneering work of Meyer (1985) it has been known that the abundances of elements in large "gradual" solar energetic particle (SEP) events are closely related to the corresponding element abundances in the solar corona. SEPs in gradual events are accelerated, in proportion to the ambient "seed population", by shock waves, driven out from the Sun by fast wide coronal mass ejections (CMEs). These drab abundances in gradual SEP events contrast sharply with the more spectacular abundances of the smaller, more numerous, "impulsive" SEP events that have 1000-fold enhancements of $^3$He/$^4$He and of heavy elements, *i.e.* ($Z{\geq}50$)/O, produced during acceleration by resonant processes in solar flares and jets (for a recent review of gradual and impulsive SEP events see Reames, 2013). The distinction of gradual and impulsive SEP events involves a wide variety of evidence (*e.g.* Reames, 1990, 1995a 1995b, 1999, 2002, 2013; Kahler, 1992, 1994, 2001; Gosling, 1993; Reames, Meyer, and von Rosenvinge 1994; Lee 1997, 2005; Mason Mazur, and Dwyer, 1999; Tylka, 2001; Gopalswamy *et al.*, 2002; Ng, Reames, and Tylka, 2003; Desai *et al.*, 2003, 2004, 2006; Slocum *et al.*, 2003; Cliver, Kahler, and Reames, 2004; Tylka *et al.*, 2005; Tylka and Lee, 2006; Cliver and Ling, 2007, 2009; Cohen *et al.*, 2007; Leske *et al.*, 2007; Ng and Reames, 2008; Sandroos and Vainio, 2009; Rouillard *et al.*, 2011, 2012; Wang *et al.*, 2012). The heavy-element enhancements, and their dependence on the mass-to-charge ratio $A/Q$ of the SEP ions, have been recently linked theoretically to the physics of magnetic reconnection regions from which the ions escape (Drake *et al.*, 2009; Knizhnik, Swisdak, and Drake, 2011; Drake and Swisdak, 2012)

Regarding the solar associations of gradual and impulsive SEP events, Kahler *et al.* (1984) established a clear association of the gradual events with fast, wide CMEs. However, Kahler, Reames, and Sheeley (2001) showed that some classic impulsive SEP events were associated with narrow CMEs and Yashiro *et al.* (2004a) found that at least 28-39% of impulsive SEPs had CME associations. These associations have been tied to the theory of jets (Shimojo and Shibata, 2000) where emerging magnetic flux reconnects on open field lines allowing SEPs and plasma, the CME, to escape (Shimojo and Shibata, 2000; Kahler, Reames, and Sheeley, 2001; Reames, 2002; Moore *et al.*, 2010; Archontis and Hood, 2013). What are the properties of the CMEs associated with the impulsive events and do they relate to the unusual abundances seen in these events?

In this paper we study a large sample (111) of Fe-rich impulsive SEP events measured by the *Low Energy Matrix Telescope* (LEMT; von Rosenvinge *et al.*, 1995) on the *Wind* spacecraft between 3 November 1994 and 5 August 2013. In Section 2 we discuss SEP event selection and the association of CMEs. Section 3 summarizes correlations among different abundances, shows that energy variations are limited, and presents the abundance enhancements from He to Pb, averaged over all of the events, *vs. A/Q.* Section 4 shows the properties of the associated CMEs, discusses possible SEP-CME correlations, characterizes C-poor and He-poor SEP events, and examines the dependence of abundance enhancements on CME speed and width. Finally, in Section 5 we discuss various aspects of the results and summarize the conclusions.

The LEMT measures elements from He through about Pb in the energy region about $2 - 20$ MeV amu$^{-1}$, identifying and binning the major elements from He to Fe onboard at a rate up to about $10^4$ particles s$^{-1}$. Instrument resolution and aspects of the processing have been shown and described elsewhere (Reames *et al.*, 1997; Reames, Ng, and Berdichevsky, 2001; Reames, 2000; Reames and Ng, 2004). Element resolution as a function of energy has been shown recently and abundances in gradual events have been analyzed by Reames (2014) using LEMT data.

## 2. Event Selection

A common description of impulsive SEP events is "Fe-rich" events. Previous studies of $^3$He-rich events have shown them to be enhanced in Fe/O by a factor of 8-10 relative to gradual events and coronal abundances (*e.g.* Mason *et al.*, 1986; Reames, Meyer, and von Rosenvinge 1994). Since many $^3$He-rich events are too small to have measurable intensities of elements with $Z$>2, they would contribute little to an abundance study, so we begin our study with a sample of Fe-rich events.

Figure 1 shows the distribution of all 8-h intervals in our 19-year period where Ne/O and Fe/O, at the lowest LEMT energies, exceed their nominal coronal values. Time periods to the right of the dashed line in Figure 1 were searched for candidate Fe-rich events for our study. Ne/O was not considered in the event selection and Fe/O is not constrained above 3 MeV amu$^{-1}$, only at its lowest energy.

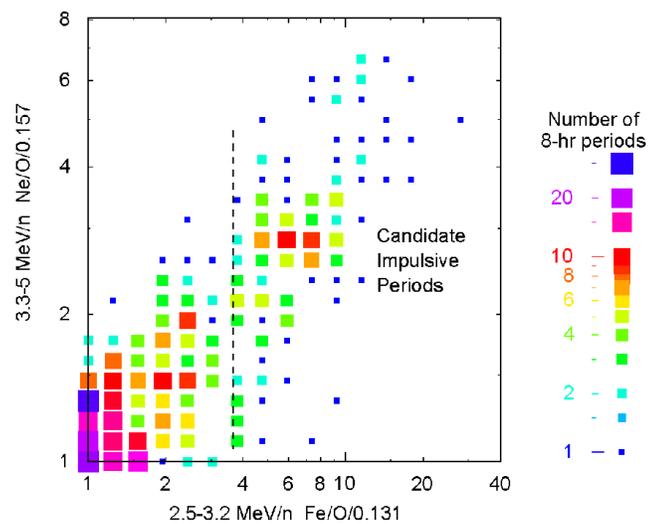

**Figure 1**. Values of enhancements of Ne/O *vs.* Fe/O are binned for all 8-h intervals which have errors of 20% or less. The cluster of periods near the origin represents gradual event periods (Reames 2014). The peak at elevated Ne/O and Fe/O is produced by impulsive events of interest in our study.

Individual SEP events were identified from intensity-time plots during these likely Fe-rich time periods and onset times were determined using 15-min averaged data. Occasionally, a new event was signaled by a sudden change in abundances (see examples in Reames and Ng, 2004). Using well-defined onsets, the associated CMEs were sought from the *Large Angle and Spectrometric Coronagraph* (LASCO; Brueckner *et al.*, 1995) on board the *Solar and Heliospheric Observatory* (SOHO), as described in Section 4.1. In Figure 2 we show some sample events with their CME associations for two groups of three SEP events. Such multiple sequential events are a common occurrence.

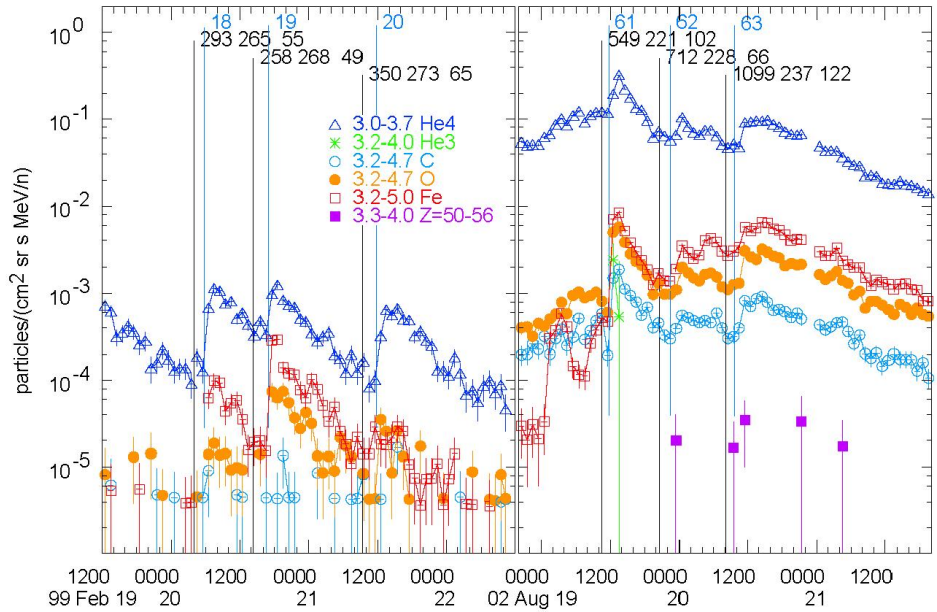

**Figure 2** Time evolution of typical Fe-rich SEP events are shown. Event onset times are flagged by the (blue) event numbers. CME onset times are flagged by three (black) numbers: the CME speed (km s$^{-1}$), the central position angle (deg) and the CME width (deg). The timing of the SEP event sequences and of the corresponding CMEs usually matches extremely well.

A complete list of our 111 impulsive SEP events with the SEP properties and the associated CMEs and other properties is included in the Appendix. In the next sections we present 1) the abundances of the events, 2) the properties of the associated CME and 3) SEP-CME relationships.

# 3. Impulsive SEP Event Abundances

### 3.1 Abundance Correlations

We begin by examining abundances in the nominal energy region of 3– 5 MeV amu[-1] where the typical relationships of the abundances are clearly shown. Figure 3 shows possible correlations of enhancements in He/O, C/O, Ne/O and Si/O with that of Fe/O. By He we always mean $^4$He throughout this paper unless the isotope is otherwise explicitly specified.

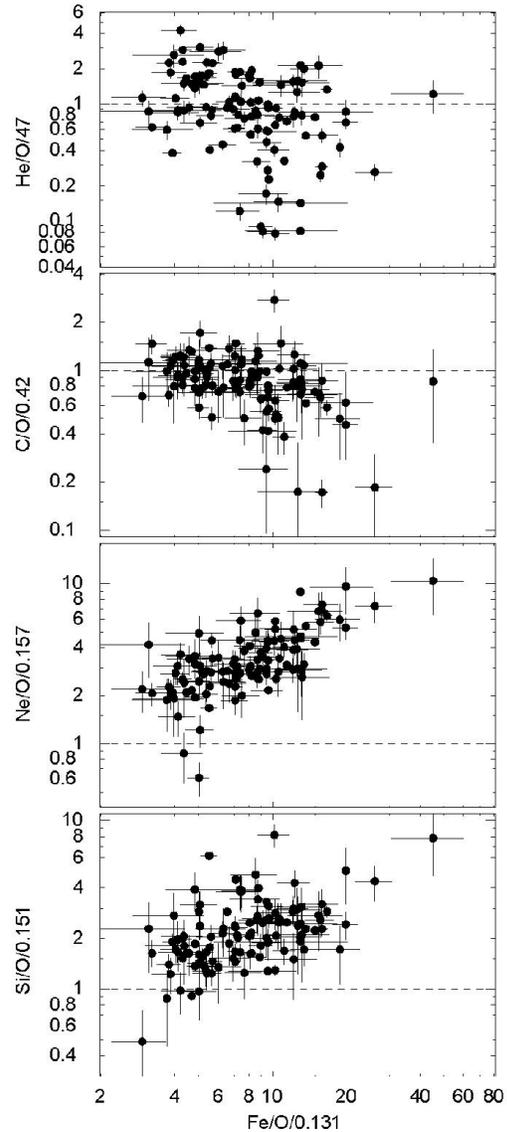

**Figure 3.** Abundances of He/O, C/O, Ne/O, and Si/O, normalized to coronal (gradual event) abundances (Reames, 2014) are plotted *vs.* the corresponding abundance of Fe/O for the 111 complete SEP events in our study, all in the nominal 3–5 MeV amu[-1] region. Ne/O and Si/O show systematic abundance displacement. Enhancements (above the dashed line) in Ne/O exceed those in Si/O. He/O and C/O have little systematic displacement from coronal values. Ne/O shows a strong positive correlation with Fe/O (correlation coefficient 0.77) while He/O shows a negative correlation coefficient of -0.55. The coefficient for C/O is -0.49 and that for Si/O is 0.44. Other energy regions show similar behavior.

The pattern of enhancements seen for Ne and Si follow behavior seen for many years (*e.g.* Reames, Meyer, and von Rosenvinge, 1994). We will pursue the more complex behavior of He and C in a Section 4.3 of this paper.

Heavy elements with atomic number $Z \geq 50$ seem to have particular relevance to the physical processes of element enhancements and have much greater leverage in abundances *vs. Z*, or *vs.* the mass-to-charge ratio *A/Q*. The unusual nature of some enhancements is difficult to exaggerate; for example, in one event the intensity of $Z \geq 50$ ions is comparable to that of C!

In Figure 4 we study the enhancement in ($Z \geq 50$) ions. The lower panel shows that enhancements are clear in high-fluence events but small events are more ambiguous. The upper panel shows that ($Z \geq 50$) ion measurability depends more upon event size (fluence) than upon Fe/O. The O fluence is the time integral of the O intensity over the duration (listed in Table A1) that that event is actually observed significantly above background.

**Figure 4.** The lower panel shows the enhancement of ($50 \leq Z \leq 92$) as a function of event size measured by the fluence of O (blue filled circles). Events with no such ions observed are plotted as yellow circles along the axis with errors showing the one-particle limit. The upper panel shows the amount of ($50 \leq Z \leq 92$) enhancement as a colored circle in a plot of Fe enhancement *vs.* O fluence.

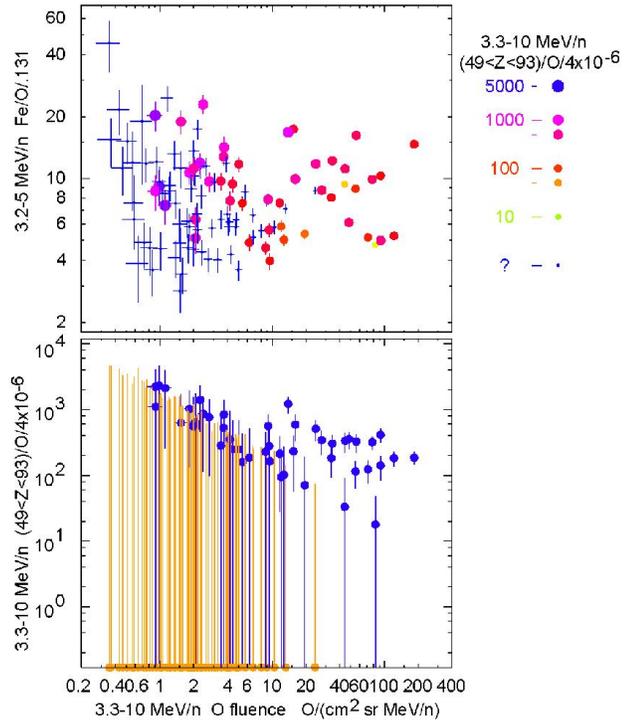

It is interesting that there is so little dependence of ($Z \geq 50$)/O upon Fe/O for impulsive events, as was also found previously (Reames and Ng, 2004). Ne/O and Si/O are more strongly correlated with Fe/O than is ($Z \geq 50$)/O.

### 3.2 Energy Dependence

It is important to know something about the energy dependence of abundances in order to assess the generality of the observations we are making about SEP abundances. Thus we consider the stability of the mean abundance values over the 2 – 15 MeV amu$^{-1}$ energy range available from LEMT.

The general behavior of the energy spectra of O is shown by simply plotting all the event O spectra using different symbols and colors in the upper panel of Figure 5. The lower panel compares spectral indices of Fe and O. Here, the tendency of points to lie above the diagonal suggests a weak bias for Fe/O to rise with energy. However, we note, additionally, that there is no correlation of the Fe spectral index with Fe/O as there was in gradual events (Reames, 2014) indicating no coupling between particle acceleration and transport.

**Figure 5** The upper panel is a composite of O energy spectra for all 111 events to show the approximate power-law spectral behavior. The lower panel shows the power-law energy spectral index of Fe *vs.* that of O for each event with the color and size of each circle showing the speed of the associated CME.

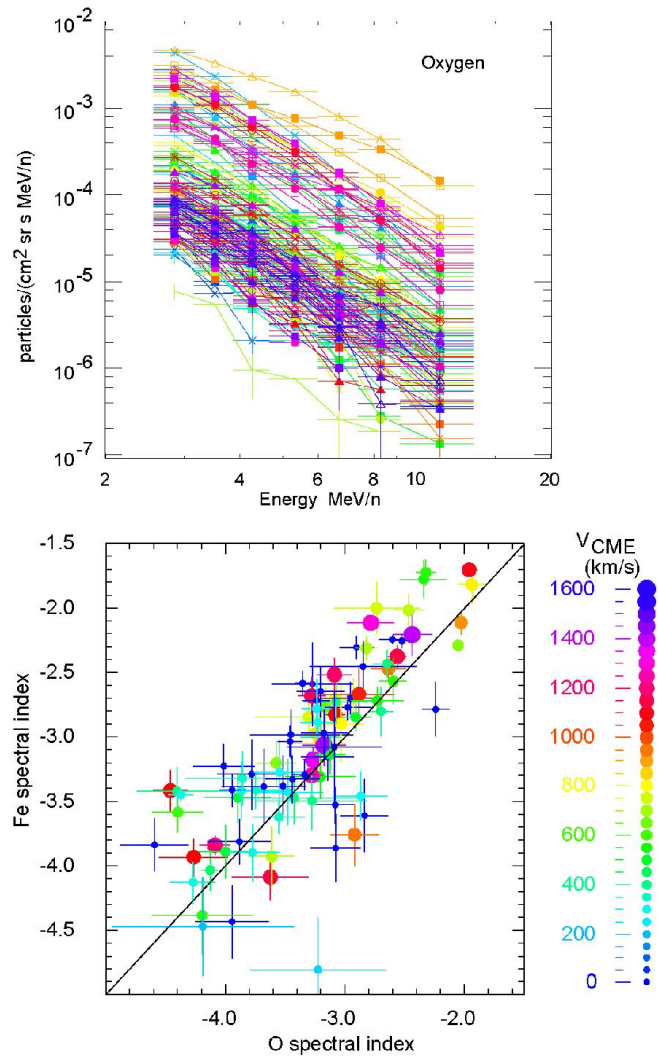

In the lower panel of Figure 5, there is a tendency for faster (and wider) CMEs to accompany harder SEP spectra. This is consistent with the model of solar jets (Shimojo and Shibata, 2000; Moore *et al.*, 2010; Archontis and Hood, 2013) because a larger energy deposit from reconnection could cause greater energization of SEPs and also drive out a larger, faster CME.

A search for possible energy dependence may be made by determining the principal abundances at several values of $E$, as was done for gradual events (Reames, 2014). For the gradual events, statistical errors were negligible so the events could be weighted equally. Here, events with abundances based upon only a few ions should not be over-weighted and an accuracy of ≈20% or better is required to clearly distinguish some enhancements. In our averaging, for weighting, we have compromised by assuming that event abundances have a fixed error of 15% in addition to any statistical error from the number of ions. This ensures that larger well-measured events are weighted equally since abundance variations among these events do not arise solely from ion counting statistics.

Figure 6 shows the abundance determined in each energy interval for each of the dominant species. There are no strong systematic variations with energy although Si and Fe may show a slight tendency to increase with $E$.

**Figure 6.** Relative abundances of different elements are determined in several energy intervals (black circles). Averages over energy for each species are shown as red squares.

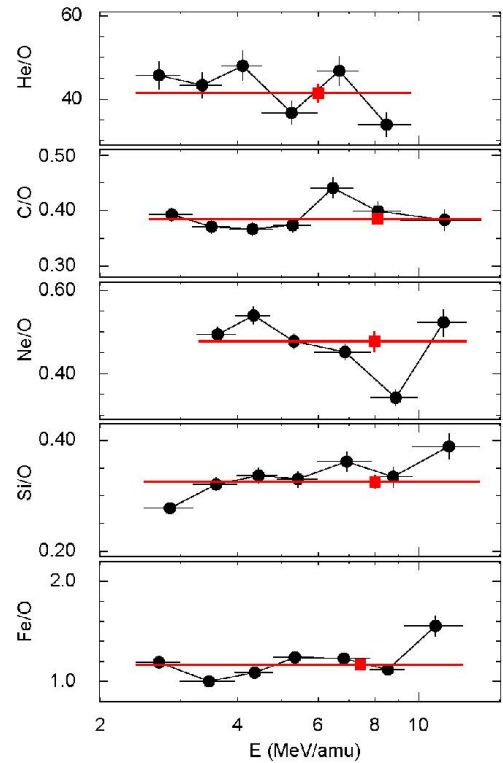

### 3.3 Average Abundances

Since Figure 6 shows that the abundances are relatively stable with energy we can supplement the abundances of major elements shown in that figure with those for species that are only well resolved over a more-limited, higher-energy regime, *e.g.* Ar and Ca, and produce the abundance comparison given in Table 1. The trans-Fe element abundances for gradual events are taken from the observed sum of ions during gradual event periods from Reames and Ng (2004). Trans-Fe abundances for the impulsive events are averages over those impulsive events for which finite abundances exist. This may cause a bias since high-abundance events are included while low-abundance events are not, for both gradual and impulsive events. However, Figure 4 suggests that all existing abundances are measureable, at least in the high-fluence region.

Table 1 Abundances in gradual and impulsive events.

| | $Z$ | Gradual SEP events[1,2] | Impulsive SEP events (this work) | Average enhancement | Nominal $A/Q$ |
|---|---|---|---|---|---|
| He | 2 | 47000±3000 | 41400±2200 | 0.88 | 2 |
| C | 6 | 420±10 | 386±8 | 0.92 | 2 |
| N | 7 | 128±8 | 139±4 | 1.09 | 2.1 |
| O | 8 | 1000±10 | 1000±10 | 1.00 | 2.2 |
| Ne | 10 | 157±10 | 478±24 | 3.04 | 2.5 |
| Mg | 12 | 178±4 | 404±30 | 2.3 | 2.4 |
| Si | 14 | 151±4 | 325±12 | 2.15 | 2.4 |
| S | 16 | 25±2 | 84±4 | 3.4 | 2.6 |
| Ar | 18 | 4.3±0.4 | 34±2 | 7.9 | 3.0 |
| Ca | 20 | 11±1 | 85±4 | 7.7 | 3.3 |
| Fe | 26 | 131±6 | 1170±48 | 8.9 | 3.7 |
| | 34-40 | 0.04±0.01 | 2.0±0.2 | 50 | 6-7 |
| | 50-56 | $(6.6±1.0)×10^{-3}$ | 2.0±0.2 | 300 | 8-9 |
| | 76-82 | $(7±3)×10^{-4}$ | 0.64±0.12 | 900 | 12-14 |

[1] Reames (2014) for $Z<30$

[2] Reames and Ng (2004) for $Z>30$

The value of $A/Q$ in Table 1 follows the logic of Reames, Meyer, and von Rosenvinge (1994) in determining $Q$, assuming all ions are at a coronal temperature in the range 2.5–3.0 MK. We have plotted $A/Q$ as a function of equilibrium coronal temperature in Figure 7.

# Abundance Enhancements in Impulsive SEP Events

As previously deduced by Reames, Meyer, and von Rosenvinge (1994), in the temperature range of interest, the elements He, C, N, and O are fully ionized, or nearly so, the elements Ne, Mg, and Si pass through states with only two orbital electrons and $Q_{Fe} \approx 15$. This region is shaded red in Figure 7. We will find additional support for this selection in Sections 4.3 and 4.4.

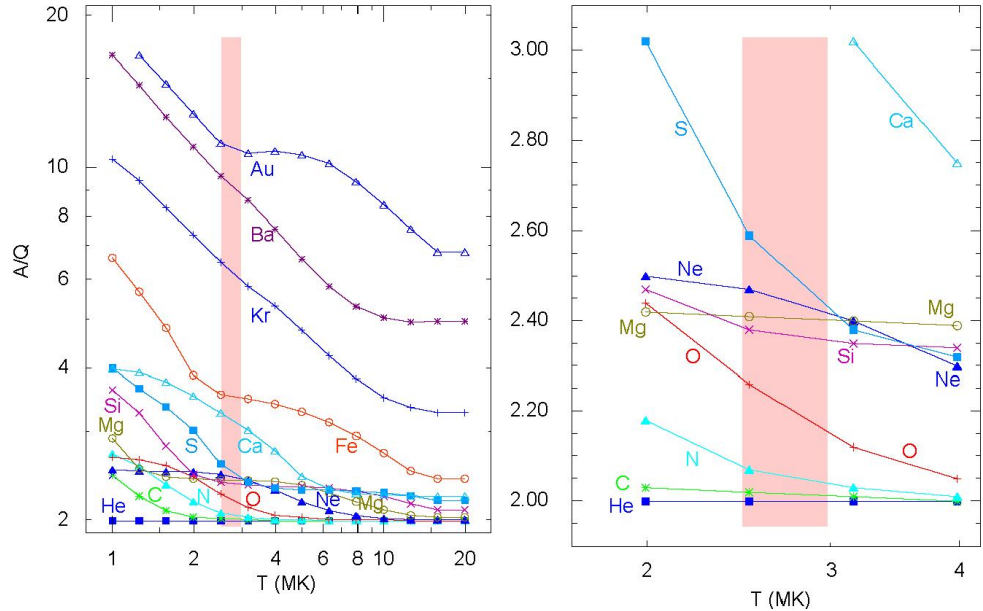

**Figure 7.** $A/Q$ is shown as a function of equilibrium temperature for several elements (left panel) and enlarged for low $Z$ (right panel). Elements below Fe are from Arnaud and Rothenflug (1985), Fe from Arnaud and Raymond (1993) and sample elements in the high-$Z$ region from Post *et al.* (1977). The region used for the likely temperature for Fe-rich impulsive SEP events is shaded red.

A plot of the mean enhancement *vs.* $A/Q$, using the values from Table 1, is shown in Figure 8. In the coronal temperature near 2.5–3 MK, $A/Q$ of Ne is actually larger than that of Mg or Si, consistent with a greater enhancement.

**Figure 8.** The mean enhancement in the abundances of elements in impulsive SEP events relative to gradual SEP events and the solar corona is shown as a function of $A/Q$ of the element. A least-squares fit line is included in the figure where the enhancement increases as the 3.64±0.15 power of $A/Q$. Both impulsive and reference abundances are shown in Table 1.

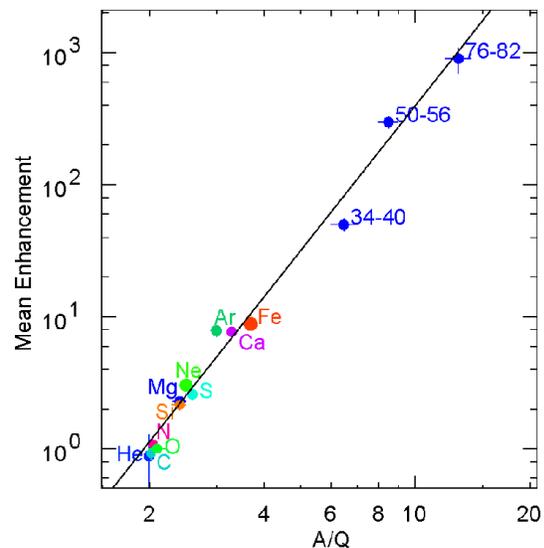

Basing the values of $Q$ upon a single equilibrium coronal temperature is clearly an approximation. The magnetic reconnection region driving CMEs from the Sun is not likely to be at equilibrium. However, no other temperature models are readily available. Also, lower ionization states of Fe at 0.06–0.11 MeV amu$^{-1}$ have been used to argue for lower source temperatures (DiFabio *et al.*, 2008, but see also Luhn *et al.*, 1984). However, ions of such different energy may come from different spatial regions in the ejecta and pickup or stripping of electrons after ion acceleration may be an important factor in producing the complex equilibrium ionization states that vary quite strongly with energy as observed.

In any case, the power-law fit shown in Figure 8 represents the data rather well. The enhancement increases as the 3.6 power of $A/Q$ for this fit. This form of dependence is consistent with the magnetic reconnection theory of Drake *et al.* (2009), Knizhnik, Swisdak, and Drake (2011) and Drake and Swisdak (2012).

## 4. CME Associations

### 4.1 Associated CME Properties

We searched the CDAW LASCO CME catalog (Yashiro *et al.*, 2004b; Gopalswamy *et al.*, 2009; http://cdaw.gsfc.nasa.gov/CME_list/) for CME associations with the 111 Fe/O-rich SEP events. Our procedure for finding associated CMEs was similar to that of Nitta *et al.*, (2006) in that we first searched the Wind/Waves low-frequency (20 kHz – 13.8 MHz) radio data (Bougeret *et al.*, 1995; http://www-lep.gsfc.nasa.gov/waves/) for candidate type III bursts for times from 1 h to 8 h before SEP event onset. We found reasonable candidate type IIIs for 95 of the 111 impulsive SEP events. In identifying these type IIIs, we gave preference to those associated with *Wind*/3DP 1 to 300 keV solar electron events for the years 1995-2005 (Wang *et al.* 2012). Of the 81 type III candidates we identified from 1995-2005, 59 (73%) had an associated electron event. The type IIIs led the electron event onsets by a median value of 5 minutes (with a range from 31 minutes before electron onset to 5 minutes after). In making the CME associations we referred to those made by Wang *et al.* (2012) for electron events and Wang, Pick, and Mason (2006) and Nitta *et al.* (2006) for subsets of the impulsive SEP events. We did not attempt to identify contributing sources to the SEP events, only the best candidate event prior to SEP onset. The listed flare locations were taken from Nitta *et al.* (2006; SXR/EIT), Wang *et al.* (2012; Hα),

*Solar-Geophysical Data* (Hα), the weekly Preliminary Reports from the Space Weather Prediction Center (Hα), and the PHTX plots in the LASCO CME catalog (Hα/EUV).  DH type II associations for the SEP events were obtained from the list on the Waves website.  CME associations were made independently by two of us (EC and SK) and several problem associations resolved.

We found 66 CME associations for the 96 events (69%) with accompanying LASCO data coverage; their speeds, central position angles, and widths are given in the Appendix.  Of these, 66 CMEs, 65 had speed determinations.  The 69% CME association we find is higher than the 28-39% rate reported by Yashiro *et al.* (2004a) for a sample of 36 impulsive events.   However, our requirement for measurable O and Fe intensities selected larger events than those in the Yashiro *et al.* (2004a) study which included small $^3$He-rich events without $Z>2$ ions.  Yashiro et al. noted that their CME association rate could be as high as 53-64% if they included 6 new faint CMEs and "obscure brightness changes" observed for 3 other events.  In a study of impulsive SEP events from 1995-2002, Nitta *et al.* (2006) reported a CME association rate of 61% (46/75).   The distributions of various properties of the 68 associated CMEs in our sample are shown in Figure 9.

D. V. Reames, E. W. Cliver, S. W. Kahler

**Figure 9** The lower panel shows the distribution of the delay between the onset of the CME and the onset of the SEP event at 3–5 MeV amu$^{-1}$. The distribution is consistent with a transport time of at least 1.5–2 h for these ions traveling a distance of 1.2 AU. The median delay of SEP onset from CME launch is 2.7 h. This is slightly longer than the 2.3 h delay measured from type III onset (due to the linear extrapolation of the height-time curve used to determine CME onset). The second panel shows the distribution of CME speeds. The median speed is 597 km s$^{-1}$ *vs.* 408 km s$^{-1}$ for all CMEs and 1336 km s$^{-1}$ for gradual SEP events (Yashiro *et al.*, 2004a). The third panel shows the distribution of CME angular width, excluding 7 halo CMEs. These CMEs are characterized as "narrow" because their median width is 75° compared to the >130° widths of CMEs associated with gradual SEP events. At least some of the CMEs with broad widths in this histogram may represent more energetic "blowout" type jets (Moore *et al.,* 2010; Archontis and Hood, 2013). The upper panel shows the longitude distribution of the associated flares. The four flares nearest central meridian are associated with halo CMEs.

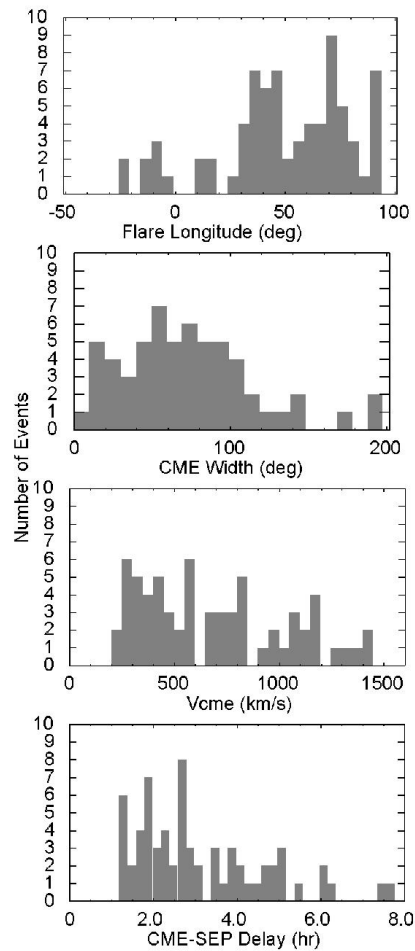

Nearly all (61/66) of the events show central position angles of the CMEs between 180° and 360°, indicating that the sources are in the western hemisphere of the Sun. 82% (54/66) of the CMEs were accompanied by SXR/EUV brightenings and/or Hα flares. Most of these flares are well-connected, as shown in the upper panel of Figure 9.

### 4.2 CME-SEP Correlations

An obvious possible relationship between impulsive SEP events and associated CMEs is the relationship of SEP intensity or fluence *vs.* CME speed that is shown for gradual SEP events by Kahler (2001). This relationship is weaker for impulsive SEP events (other differences between gradual and impulsive events have been reviewed by Reames (2013)). Note that the CME speeds measured by LASCO are speeds that are projected onto the plane of the sky.

The upper panel of Figure 10 shows the O fluence, a measure of event size, *vs.* CME speed. The unweighted correlation coefficient for these data is only 0.41, a moderately poor correlation. The correlation is not improved significantly by either considering or removing wide CMEs or those with dekametric/hectometric (DH) type II radio bursts often accompanying gradual events.

**Figure 10.** The upper panel shows the O fluence in each event *vs.* the CME speed. Color and size of the symbols indicate the CME width and events with DH type II radio emission are circled. The lower panel re-orients the same data by displaying O fluence as a colored symbol on a plot of CME width *vs.* CME speed. Again events with DH type II are circled. Events without type II are usually narrow.

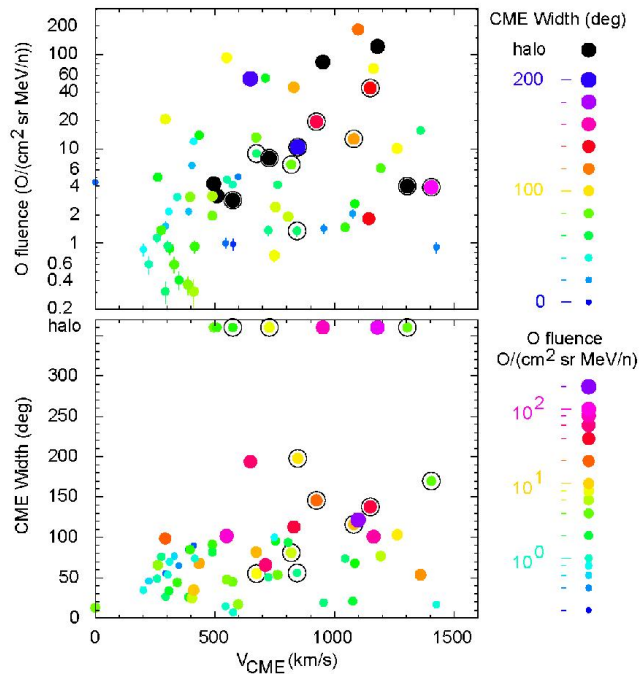

The lower panel in Figure 10 stresses the narrowness of CMEs in most impulsive events. If we exclude halo events or those with DH type II bursts, all but one of the remaining events has a width ≤100°. However, lower-fluence events tend to be much narrower than high-fluence events. In fact, most events with width <50° have quite low O fluences. Fluences appear to depend almost as strongly upon CME width as upon CME speed.

For these narrow CMEs, the efficiency of the connection of the observer may be as important in determining the SEP event fluence (*i.e.* intensity or duration) as is the intensity within the magnetic flux tube connected to the CME source.

### 4.3 He-poor and C-poor Events

There was early evidence of "C-poor" periods, with low C/O, accompanying Fe-rich events (Mason, Gloeckler, and Hovestadt, 1979) but subsequent studies found such events to be relatively rare (*e.g.* Mason *et al.*, 1986, Reames, Meyer, and von Rosenvinge, 1994). Figure 3 shows some events with low C/O and, in addition, low He/O spreading below the bulk of the distributions. Note that these events are also quite Fe-rich.

In the lower two panels of Figure 11 we show C/O and He/O *vs.* CME speed. C-poor and He-poor events occur for slow (<700 km s⁻¹) CMEs or for events with no visible CME, possibly also indicating a slow, weak CME that is difficult to see. The upper panel shows a peak near coronal abundances for all events with fast CMEs and many events with slow on no CMEs. Events that are C-poor, He-poor, or both, are all associated with slow or unseen CMEs.

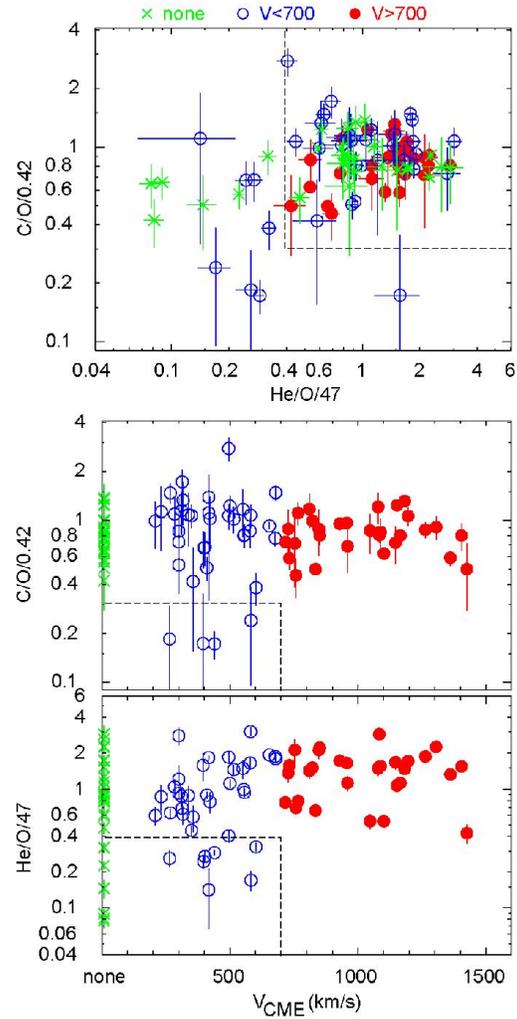

**Figure 11.** The middle and lower panels show relative abundances of C/O and He/O as a function of CME speed. Dashed lines enclose very C-poor and He-poor events, respectively, at CME speeds below 700 km s⁻¹. CME speeds are identified as fast (>700 km s⁻¹, red solid circle), slow (<700 km s⁻¹, blue open circle) and no CME seen (green cross). The upper panel shows C/O *vs.* He/O with CME speed indicated and the dashed lines bounding the C- and He- poor region to the lower left. All fast CMEs and many others contribute to a peak near normal coronal abundances in the upper panel.

To understand these observations, we assume that all events involve fully ionized He with $(A/Q)_{He} = 2$; enhancements of other partially-ionized species can increase as a function of $A/Q$ of the species for $A/Q > 2$. In hot plasma, when C and O are fully ionized, they also have $A/Q = 2$ and must be unenhanced relative to He (see Figure 7). In this case, the abundances of He, C, and O (and also N) will all have the same coronal abundances as also seen in gradual SEP events. In less energetic events with slow CMEs and cooler plasma, O (and possibly C) can be partially ionized with $A/Q > 2$ so that O is enhanced relative to He, appearing as a He-poor event. Events with O enhanced, but C and He unenhanced will be both He-poor and C-poor. Events with both C and O enhanced may only be He-poor. Only one event in Figure 11 appears to be C-poor, by one standard deviation, but not He-poor; if statistically valid this one event would not be explained by this model.

The abundance variations we have described are often quite evident in the intensity-time profiles of the events. We show these for a sample of events in Figure 12.

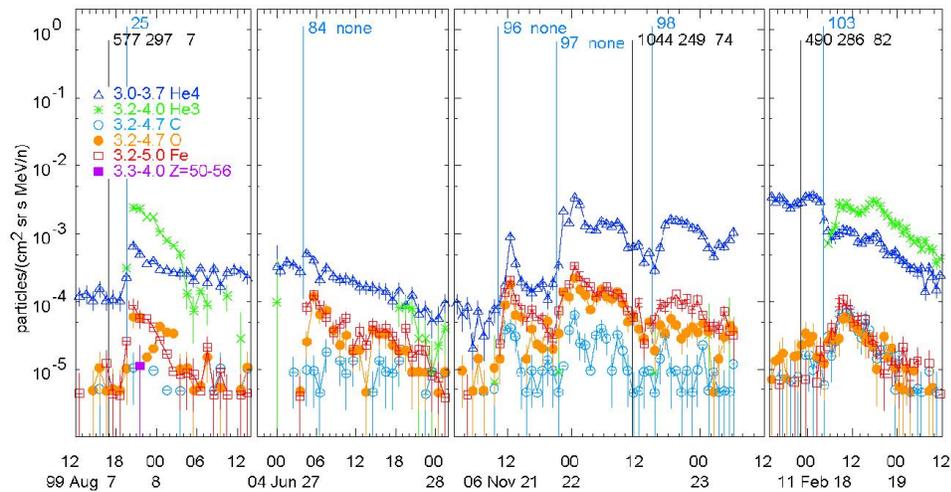

**Figure 12.** Intensity-time plots are shown for sample events showing clear abundance differences (see text). Event onsets are flagged with (blue) event numbers and CME onsets with (black) CME parameters as described in Figure 2.

Event 25 in August 1999 in Figure 12 is strongly $^3$He-rich ($^3$He/$^4$He=3.7±0.3). It is He-poor ($^4$He/O=8.0±1.6) and C-poor (C/O=10±6%) and the intensity of 50≤Z≤56 ions is comparable with that of C. Event 84 is extremely He-poor (He/O=3.8±0.4) and C-poor (18±5%). Event 96 is He-poor (3.6±0.1) but subsequent events are not. Event 103 is a unique event in our list in being *C-rich* (C/O=1.16±0.18) and somewhat He-poor (He/O=19±2). This could occur in the unlikely event that $A/Q$ was larger for C than for O or if the background plasma already had C/O ≈ 1 initially – involving cool plasma (and low CME speed) in any case.

### 4.4 Abundance Variations and CMEs

In Section 3.3, we discussed the dependence of the average abundance enhancements upon $A/Q$ and the dependence of $A/Q$ upon coronal temperature, but how can we understand the broad and correlated behavior of the event-to-event abundance variations shown in Figure 3, for example. The strong correlated behavior can not result from statistical fluctuations. We take our cue from the apparent temperature dependence in the He-poor events (Section 4.3) since $A/Q$ is always 2 for He for these events. Figure 13 shows the enhancements of typical elements, normalized to He, and plotted as a function of $Z$, for events with the indicated CME width or speed. To limit the number of events and improve visibility, we have only plotted those that have a non-zero value of $(Z\geq50)/O$.

**Figure 13.** Enhancements of specific elements relative to He are shown as a function of $Z$ for individual SEP events with a non-zero value of $(Z\geq50)/O$. The circle size and color plotted for each event depend upon the CME speed (lower panel) or width (upper panel). Some He-poor events are identified at O. An increase in energy from reconnection causes more plasma heating that affects $A/Q$ and hence the pattern of enhancements; $A/Q$ can vary for each of the elements with $Z>2$ (see text). The local rise at Ne occurs because the $A/Q$ at Ne is larger than at either O or Si (see Figure 7).

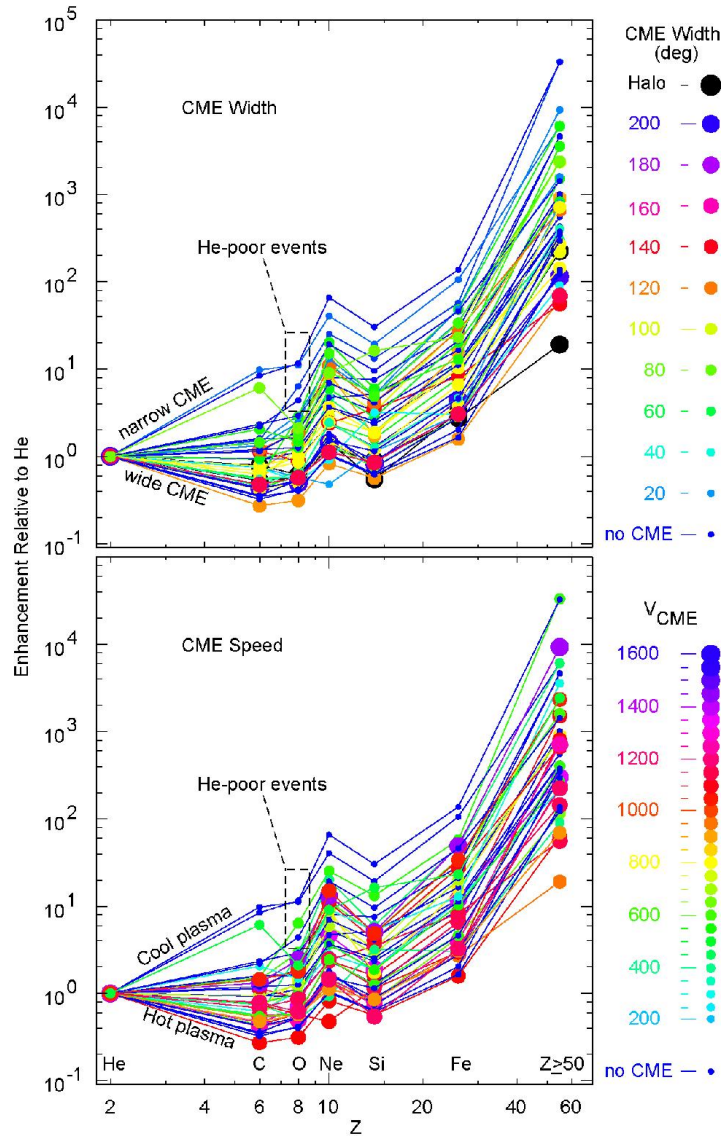

It is clear in Figure 13 that for each element with $Z$>2, the small blue or green circles indicating CME speeds <700 km s$^{-1}$ (lower panel) or widths <100$^o$ (upper panel), tend to fall above the large red or violet circles where speeds are >700 km s$^{-1}$ or widths >100$^o$.  Of course, statistical variations are not absent, for example the enhancements of C and O cannot really be below that of He, since $A/Q$ must be ≥2.  However, it is clear that events with faster, wider CMEs lie along the bottom of the distribution where the enhancements of C and O are smaller, as would be expected for smaller $A/Q$ seen at T≥3–4 MK (see Figure 7) where O is nearly fully ionized.  Events with narrow, slow CMEs show rising enhancements at O, relative to He and C (typical of rising $A/Q$ near and below 2.5 MK) and eventually blend into the cool He-poor events.  It is only when we normalize the enhancements to He, magnifying the element ionization differences, that the dependence on CME speed begins to emerge.

Events with the narrowest, slowest CMEs tend to have the greatest SEP abundance enhancements while those with wider, faster CMEs have more modest enhancements.  Thus it is possible to understand all of the observed abundance variations in impulsive SEP events in terms of coronal energy deposit from magnetic reconnection that, 1) raises temperatures increasing ionization and lowering $A$/Q in the SEP acceleration region, and 2) drives CME speed and size.  Element enhancements are controlled by their power-law dependence of upon $A/Q$ as seen in Figure 8.   However, we cannot be certain that the power law in individual events is identical to that shown for the average of all events in Figure 8.

If we correlate the logarithms of the element abundances C/He, O/He, Ne/He, Si/He, Fe/He and ($Z$≥50)/He *vs.* CME width, we find the correlation coefficients (unweighted): -0.30, -0.33, -0.28, -0.37, -0.21, and -0.19, respectively.  Note, however, that the CME speed and width are correlated with a coefficient of 0.34 for our events.

 Since we also found some correlation of the O fluence with CME speed and width, it seems reasonable to see how CME width falls on a plot of Fe/He (an abundance poorly correlated above), *vs.* fluence, as shown in Figure 14.

**Figure 14.** A plot is shown of the enhancement of Fe/He above the coronal value (dashed line) as a function of the O fluence for events, with symbol size and color indicating the width of the associated CME.

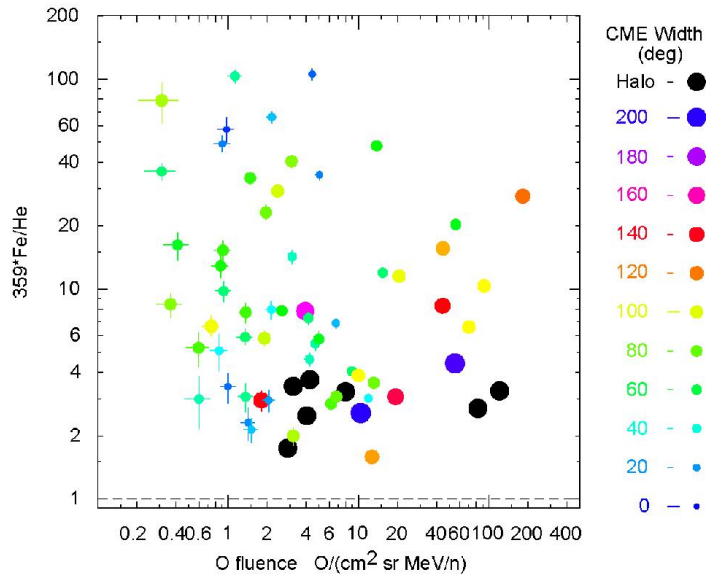

Clearly the wide CMEs and halo CMEs strongly tend to cluster in the lower right corner of Figure 14, while the narrow CMEs cluster toward the left and dominate the large Fe/He enhancements. The effects of both abundance and SEP intensity (fluence) can be seen. The effect of CME width on Fe/He abundance is clear despite the presence of 2 or 3 wider CMEs that reduce the correlation of Fe/He *vs.* width. Other abundances show very similar behavior.

## 5. Discussion and Conclusions

In Section 4.4 we have understood the fact that there is a large region of correlated variation of Ne, Mg, Si, and Fe (see Figure 3) over a factor of $\approx$4–5, slightly larger for heavier species. The events in Ne/O *vs.* Fe/O (Figure 3) are *not* in a Gaussian distribution; there is a spread that we have understood by temperature variations in the magnetically reconnecting SEP source plasma. The fact that the enhancement of Ne/O exceeds that of Si/O for most events suggests an equilibrium temperature in the range of 1.5–3.0 MK where $A/Q$ for Ne exceeds that of Mg and Si (see Figure 7) while the high ionization of C, N, and O suggests a temperature near the top of this range. In fact, in the region of 2.5–3 MK, $A/Q$ for Ne, Mg, and Si vary inversely with $Z$ (Figure 7) and so do the corresponding enhancements (see Table 1). Thus the abundances suggest that the temperature changes and the event-to-event variations depend *strongly* upon $A/Q$, as well as possible variations in the strength of the acceleration (*e.g.* Drake *et al.* 2009).

If we compare SEP properties with soft X-ray intensities, for the associated flares, on the GOES logarithmic scale (where B, C, M, and X represent intensities above $10^{-7}$, $10^{-6}$, $10^{-5}$, and $10^{-4}$ W m$^{-2}$, respectively), we find generally poorer correlations than we found with CMEs. O fluence *vs.* X-ray intensity has a correlation coefficient of 0.3, but Fe/O and even Fe/He, with coefficients 0.09 and -0.09, respectively, are uncorrelated with intensity. The He-poor events are somewhat of an exception since they are all associated with only B- and C-class flares, while the other SEP-associated flares span the B-, C-, M-, and even X-class flares (mostly C- and M-classes). It is also true that the 17 events with Fe/He enhancements > 30 are associated with only B- and C-class flares (compare Figure 14). More generally, it would not be surprising that SEPs and flares are poorly correlated if SEPs are accelerated at reconnection sites on open field lines while flares involve reconnection on neighboring closed loops.

It is possible to derive an X-ray temperature by fitting two soft X-ray channels to a thermal distribution; such temperatures have been compared for SEP proton events (Garcia 1994). However, these X-ray temperatures lie in the range 10–30 MK, even for M-class events. Ne is fully ionized in this temperature range, so that the observed enhancements in Ne/O would be difficult to explain. Perhaps these X-ray temperatures are spectral parameters and not actual plasma temperatures; in addition, the SEPs and X-rays may arise from spatially different regions.

For those events where the CME can drive a shock wave, we cannot rule out additional acceleration of the SEPs by the CME-driven shock wave. Shock acceleration could reduce the abundance enhancements by incorporating ambient plasma with coronal abundances, but the pre-accelerated energetic ions from the reconnection region would be preferentially accelerated by these weak shocks. In fact, quasi-perpendicular shock waves might preferentially enhance any ions from impulsive events that are present in the seed population (see *e.g.* Tylka and Lee, 2006). Only 11 (10%) of the events have DH type II emission indicating a shock wave strong enough to persist above ~3 solar radii (Cliver, Kahler, and Reames, 2004). However, shocks may need a quasi-perpendicular region to accelerate the electrons that produce type II emission (*e.g.* Ganse *et al.*, 2012) so the total number of shocks may be somewhat higher. Nevertheless, the existence of shock acceleration in several events does not alter our conclusions about the origin of the SEP abundance enhancement in open magnetic-reconnection source regions, its dependence on $A/Q$, and the related production of CMEs.

A [3]He-rich event with unusual abundances including large enhancement of N was found by Mason *et al.* (2002). The authors interpreted this event as having a coronal temperature of <1.5 MK. Such events are very small, they are seen primarily at energies below 1 MeV amu[-1], and that event was not seen by LEMT. Such small impulsive SEP events with cooler plasma have an unusual pattern of values of $A/Q$; larger, more energetic impulsive SEP events, that we study here, usually have a more consistent pattern of $A/Q$ like that shown in Table 1 (see also the discussion of temperature in Wiedenbeck *et al.* (2010)).

In [3]He-rich impulsive events Temerin and Roth (1992; see also Roth and Temerin, 1997), found that streaming electrons produced electromagnetic ion cyclotron waves that could resonantly accelerate [3]He ions mirroring in the magnetic field, in analogy with acceleration of "ion conics" seen in the Earth's aurora. Heavier elements could be enhanced by second-harmonic interaction with these waves but this does not lead to a power-law enhancement with $A/Q$. In fact, it is clear that while impulsive events have enhancements in both [3]He and heavy ions, these enhancements are completely uncorrelated (*e.g.* Reames, Meyer, and von Rosenvinge, 1994), suggesting different mechanisms for acceleration of [3]He and heavy ions. The $A/Q$ dependence for heavy ions does seem consistent with the magnetic reconnection theory of Drake *et al.* (2009; Knizhnik, Swisdak, and Drake, 2011; Drake and Swisdak, 2012).

It is not surprising that an energetic reconnection event that results in the free escape of high energy ions into space, an impulsive SEP event, might also eject bulk plasma into space, a CME. This is, in fact, an expected consequence of the combined CME/SEP model of solar jets (Shimojo and Shibata, 2000; Kahler, Reames, and Sheeley, 2001; Reames, 2002; Moore *et al.*, 2010; Archontis and Hood, 2013) where the magnetic reconnection involves emerging flux reconnecting with open field lines (see also Nitta *et al.*, 2006; Wang, Pick, and Mason, 2006). Impulsive SEP events and CMEs appear to be fundamentally linked, even when the latter are too weak to see. The high degree of association we find between these phenomena (69%) leads us to speculate that all (impulsive) Fe-rich events have associated CMEs. However, not all CMEs may come from reconnections that are energetic enough to produce a detectible SEP event. In addition, many CMEs undoubtedly lack associated observed SEP events because they are not magnetically connected to the SEP observer

By way of contrast, we point out that gradual events are associated with fast, wide CMEs that drive strong shock waves that accelerate SEP ions of the ambient plasma of the corona and solar wind. Thus both event types are CME-associated but the CMEs and the physics are different. For impulsive SEPs, mass is ejected on open field lines as a jet, while for gradual events the CMEs result from erupting, magnetically-closed structures. For gradual events the acceleration occurs at the shock, while for impulsive events the acceleration occurs at the open magnetic reconnection region driving the ejecta.

In summary, Fe-rich impulsive SEP events are strongly (69%) associated with CMEs. The CMEs are relatively narrow (median width of 75º) and all but a few originate on the western side of the Sun.

These impulsive SEP events show a well-defined pattern of enhanced abundances of elements relative to coronal abundances that increases as a power law in *A/Q* from an average factor of 3 at Ne to a factor of 900 at Pb. The elements He, C, N, and O in the SEPs are usually nearly fully ionized in the hot, 2.5–3 MK, coronal plasma associated with the faster CMEs and maintain their coronal abundances. However, SEP events associated with slower CMEs arise from cooler plasma where elements from C to Pb have much greater enhancements and where partially-ionized O is enhanced to produce SEPs called He-poor or C-poor events.

**Acknowledgments**: S.K. and E.C. were funded by AFOSR Task 2301RDZ4. CME data were taken from the CDAW LASCO catalog. This CME catalog is generated and maintained at the CDAW Data Center by NASA and The Catholic University of America in cooperation with the Naval Research Laboratory. SOHO is a project of international cooperation between ESA and NASA. We acknowledge the use of the Wind/WAVES data.

## Appendix

Table A1 lists properties of 111 Fe-rich impulsive SEP events. Columns in the table are the event number, the SEP onset time, the SEP event duration (Dur.), the abundance ratios of Fe/O, Ne/O, and He/O at 3–4 MeV amu$^{-1}$, the ratio (50≤$Z$≤56)/O at 3–10 MeV amu$^{-1}$, the associated CME speed, $V$, the central position angle (CPA) and width ($W$) of the CME, the delay between the CME onset and the SEP onset, the UT onset time of the DH type III burst (* indicates previous day), the presence of a DH type II radio burst, and the location of associated flare or other source. NA in the CME speed indicates that no LASCO coronagraph data were available, dash indicates that data were available but no CME was listed.

**Table A1**. Properties of the Fe-rich SEP events and associated CMEs

| | SEP Onset | Dur. [h] | Fe/O | Ne/O | He/O | (50≤$Z$≤56)/O [×10$^3$] | $V$ [km s$^{-1}$] | CPA [deg] | $W$ [deg] | Delay [h] | DHIII Onset | DH II | Location |
|---|---|---|---|---|---|---|---|---|---|---|---|---|---|
| 1 | 95 Apr 2 1300 | 17 | 1.71±0.69 | 0.74±0.36 | 3.8±1.5 | 0±66 | NA | | | | 1105 | | – |
| 2 | 96 Jul 12 1715 | 34.8 | 1.60±0.23 | 0.61±0.11 | 74±9 | 9.1±6.5 | 1085 | 260 | 68 | 2.0 | 1525 | | S11W72 |
| 3 | 97 Sep 18 0215 | 15.8 | 0.92±0.09 | 0.53±0.06 | 28.8±2.2 | 1.8±1.8 | – | – | – | – | 0000 | | S25W76 |
| 4 | 97 Sep 18 2300 | 17 | 0.82±0.10 | 0.44±0.07 | 21±2 | 0±2.8 | 343 | 275 | 44 | 3.6 | 1950 | | S23W90 |
| 5 | 97 Sep 20 1200 | 22 | 1.14±0.20 | 0.53±0.12 | 37.9±5.5 | 0±4.8 | – | – | – | – | 1025 | | – |
| 6 | 97 Oct 20 0630 | 11.5 | 1.71±0.95 | 1.4±0.8 | 6.7±3.5 | 0±39 | 412 | 274 | 90 | 3.0 | 0335 | | – |
| 7 | 98 May 4 0000 | 24 | 1.08±0.04 | 0.416±0.020 | 90.7±2.4 | 0.21±0.21 | 649 | 317 | 194 | 2.7 | 2115* | | S13W34 |
| 8 | 98 May 29 0530 | 10.5 | 0.51±0.08 | 0.33±0.07 | 87±9 | 0±4.3 | 489 | 274 | 91 | 4.9 | 0100 | | N18W90 |
| 9 | 98 May 29 1815 | 8.75 | 0.52±0.17 | 0.30±0.13 | 123±26 | 0±17 | – | – | – | – | – | | – |
| 10 | 98 Aug 13 1930 | 46.5 | 1.42±0.33 | 0.71±0.20 | 68±14 | 0±13 | NA | | | | 1755 | | S25W61 |
| 11 | 98 Sep 6 0700 | 23 | 1.76±0.21 | 0.49±0.08 | 94±10 | 4.8±3.4 | NA | | | | 0550 | | – |
| 12 | 98 Sep 9 0645 | 14.3 | 1.07±0.05 | 0.475±0.029 | 25.8±0.9 | 0±0.39 | NA | | | | 0455 | | – |
| 13 | 98 Sep 26 1830 | 15.5 | 1.16±0.08 | 0.462±0.041 | 71.8±3.7 | 3.3±1.9 | NA | | | | 1620 | | N22W39 |
| 14 | 98 Sep 27 1030 | 8 | 0.73±0.03 | 0.362±0.020 | 19.0±0.6 | 2.4±0.86 | NA | | | | 0810 | | N21W48 |
| 15 | 98 Sep 27 1900 | 7 | 1.40±0.07 | 0.537±0.035 | 35.8±1.4 | 4.5±1.6 | NA | | | | 1625 | | N21W52 |
| 16 | 98 Sep 28 0230 | 17.5 | 1.50±0.06 | 0.490±0.027 | 33.4±1.1 | 1.2±0.68 | NA | | | | 2340* | | N20W58 |
| 17 | 98 Sep 29 0615 | 18.8 | 0.94±0.07 | 0.479±0.046 | 29.4±1.6 | 1.2±1.2 | NA | | | | 0200 | | N23W69 |
| 18 | 99 Feb 20 0545 | 10.3 | 5.9±1.9 | 1.64±0.63 | 57±18 | 0±38 | 293 | 265 | 55 | 1.8 | 0405 | | S18W63 |
| 19 | 99 Feb 20 1700 | 6 | 3.44±0.59 | 1.14±0.24 | 12.3±2.0 | 0±8.6 | 258 | 268 | 49 | 2.6 | 1510 | | S21W72 |

| | | | | | | | | | | | | | | |
|---|---|---|---|---|---|---|---|---|---|---|---|---|---|---|
| 20 | 99 Feb 21 | 1145 | 12.3 | 1.26±0.39 | 0.68±0.26 | 27±7 | 0±21 | 350 | 273 | 65 | 2.3 | 0945 | | S17W85 |
| 21 | 99 May 12 | 0715 | 6.75 | 0.41±0.16 | 0.66±0.25 | 40±10 | 0±29 | 224 | 265 | 46 | 1.4 | 0535 | | S20W90 |
| 22 | 99 Jun 13 | 0430 | 19.5 | 0.38±0.10 | 0.35±0.10 | 53±8 | 0±7.5 | 955 | 239 | 19 | 2.8 | 0150 | | − |
| 23 | 99 Jun 18 | 1200 | 19 | 1.07±0.12 | 0.48±0.07 | 48.5±4.3 | 0±2.2 | NA | | | | | | − |
| 24 | 99 Jul 12 | 1100 | 61 | 0.92±0.11 | 0.294±0.056 | 54.3±5.1 | 0±2.0 | − | − | − | − | − | | − |
| 25 | 99 Aug 7 | 1930 | 7.5 | 1.24±0.28 | 0.621±0.18 | 8.02±1.5 | 28±20 | 577 | 297 | 7 | 2.7 | 1705 | | N22W74 |
| 26 | 99 Sep 19 | 1630 | 15.5 | 0.66±0.12 | 0.39±0.09 | 79±10 | 0±6.1 | 1144 | 307 | 139 | 2.7 | 1410 | | N21W71 |
| 27 | 99 Nov 16 | 0830 | 27.5 | 0.66±0.07 | 0.389±0.053 | 80.4±5.8 | 0±2.0 | 1193 | 270 | 77 | 1.8 | 0610 | | N09W42 |
| 28 | 99 Dec 24 | 0430 | 16.5 | 1.26±0.18 | 0.34±0.08 | 43±5 | 0±4.6 | − | − | − | − | − | | − |
| 29 | 99 Dec 25 | 0100 | 35 | 1.61±0.24 | 0.70±0.13 | 39.7±5.0 | 0±5.3 | − | − | − | − | − | | − |
| 30 | 99 Dec 26 | 1530 | 13.5 | 2.6±0.8 | 1.51±0.49 | 40.2±10 | 0±24 | − | − | − | − | 1320 | | − |
| 31 | 99 Dec 27 | 0515 | 20.8 | 2.61±0.31 | 0.84±0.13 | 32.6±3.5 | 4.6±4.6 | 753 | 283 | 96 | 3.8 | 0135 | | N24W35 |
| 32 | 99 Dec 28 | 0430 | 31.5 | 0.93±0.06 | 0.358±0.035 | 88.2±4.5 | 0±0.6 | 672 | 293 | 82 | 4.0 | 0045 | | N20W56 |
| 33 | 00 Feb 17 | 2330 | 72.5 | 0.66±0.06 | .0963±0.023 | 74.1±5.0 | 0±1.3 | 728 | 184 | 360 | 3.4 | 2030 | Y | S29E07 |
| 34 | 00 Mar 7 | 1445 | 11.3 | 2.05±0.28 | 0.91±0.15 | 11.5±1.4 | 0±3.9 | 391 | 269 | 26 | 2.8 | 1230 | | S15W72 |
| 35 | 00 Mar 8 | 0115 | 8.75 | 1.18±0.13 | 0.58±0.08 | 4.2±0.40 | 2.3±2.3 | − | 264 | 13 | 1.3 | 2340* | | S15W76 |
| 36 | 00 Mar 19 | 1600 | 28 | 1.72±0.36 | 0.45±0.14 | 37±6 | 0±8.5 | − | − | − | − | 1245 | | S18W74 |
| 37 | 00 May 1 | 1130 | 13.5 | 2.18±0.12 | 0.99±0.06 | 62.1±2.9 | 2.1±1.2 | 1360 | 323 | 54 | 1.3 | 1020 | | N21W50 |
| 38 | 00 May 4 | 1315 | 22.8 | 1.59±0.19 | 0.46±0.08 | 73±7 | 0±2.4 | 1404 | 235 | 170 | 2.3 | 1105 | Y | S17W90 |
| 39 | 00 May 23 | 1800 | 42 | 0.95±0.04 | 0.43±0.024 | 37.9±1.2 | 1.1±0.66 | − | − | − | − | 1640 | | N22W38 |
| 40 | 00 Jun 4 | 0830 | 15.5 | 1.47±0.15 | 0.63±0.08 | 15.3±1.3 | 0±2.8 | 597 | 295 | 17 | 1.5 | 0705 | | S10W62 |
| 41 | 00 Jun 15 | 2100 | 27 | 0.566±0.046 | 0.399±0.040 | 136±7 | 0.91±0.91 | 1081 | 295 | 116 | 1.6 | 1940 | Y | N20W65 |
| 42 | 00 Jun 19 | 1000 | 10 | 0.99±0.17 | 0.31±0.09 | 67±9 | 0±5.8 | 806 | 303 | 94 | 6.1 | 0415 | | − |
| 43 | 00 Jun 23 | 1915 | 26.8 | 0.75±0.07 | 0.537±0.058 | 105±7 | 0±1.3 | 847 | 282 | 198 | 5.4 | 1420 | Y | N26W72 |
| 44 | 00 Jul 11 | 0200 | 22 | 0.74±0.07 | 0.70±0.07 | 41.6±2.8 | 0±1.5 | 404 | 238 | 25 | 1.7 | 2350* | | − |
| 45 | 00 Aug 12 | 1400 | 5 | 2.09±0.12 | 1.07±0.07 | 13.6±0.7 | 4.9±2.2 | 434 | 307 | 68 | 1.9 | 1230 | | N05W48 |
| 46 | 00 Dec 28 | 0100 | 15 | 1.14±0.15 | 0.43±0.08 | 15.1±1.6 | 4.1±4.1 | − | − | − | − | 2340* | | N13W36 |
| 47 | 01 Mar 10 | 1100 | 49 | 0.63±0.063 | 0.307±0.043 | 71.2±4.8 | 0±1.1 | 819 | 297 | 81 | 7.6 | 0405 | Y | N27W42 |
| 48 | 01 Mar 21 | 1000 | 54 | 0.58±0.18 | 0.38±0.15 | 41±9 | 0±19 | 331 | 272 | 77 | 8.1 | 0235 | | S06W64 |
| 49 | 01 Apr 14 | 1845 | 13.3 | 1.34±0.051 | 0.916±0.040 | 30.9±0.9 | 2.7±0.9 | 830 | 263 | 113 | 1.2 | 1705 | | S18W71 |
| 50 | 01 Apr 26 | 1645 | 9.25 | 1.71±0.33 | 0.72±0.18 | 100±16 | 0±8.3 | 844 | 271 | 56 | 3.8 | 1310 | Y | N17W31 |
| 51 | 01 Jun 25 | 0800 | 28 | 0.66±0.18 | 0.77±0.21 | 70±13 | 0±9.4 | 545 | 77 | 15 | 4.3 | 0350 | | − |
| 52 | 01 Sep 10 | 1645 | 34.3 | 1.35±0.07 | 0.402±0.029 | 43.3±1.7 | 0.98±0.7 | 293 | 276 | 99 | 5.0 | 1320 | | N18W90 |
| 53 | 01 Sep 22 | 0800 | 16 | 1.61±0.38 | 0.82±0.24 | 37±7 | 0±13 | 416 | 262 | 74 | 2.5 | 0540 | | S09W65 |
| 54 | 02 Feb 20 | 0715 | 28.8 | 0.586±0.020 | 0.330±0.015 | 77.4±1.8 | 0.15±0.15 | 952 | 263 | 360 | 1.3 | 0555 | | N12W72 |

| # | Date | Time | | | | | | | | | | | | | |
|---|---|---|---|---|---|---|---|---|---|---|---|---|---|---|---|
| 55 | 02 Apr 14 | 2330 | 4.5 | 0.792±0.19 | 0.54±0.16 | 132±22 | 0±8.1 | 294 | 323 | 27 | 2.0 | 2225 | | N18W75 |
| 56 | 02 Apr 15 | 0445 | 15.3 | 0.93±0.08 | 0.402±0.047 | 84.5±5.4 | 0±1.4 | 674 | 308 | 55 | 2.6 | 0250 | Y | N19W79 |
| 57 | 02 Aug 1 | 1100 | 13 | 0.874±0.15 | 0.37±0.09 | 48.9±6.2 | 4.9±4.9 | – | – | – | – | – | | – |
| 58 | 02 Aug 2 | 1800 | 22 | 0.91±0.15 | 0.450±0.11 | 42.4±5.3 | 0±4.6 | – | – | – | – | – | | – |
| 59 | 02 Aug 3 | 2345 | 15.3 | 1.15±0.046 | 0.4±0.023 | 50.4±1.6 | 0±0.28 | 1150 | 259 | 138 | 5.0 | 1910 | Y | S16W80 |
| 60 | 02 Aug 4 | 1745 | 42.3 | 1.24±0.035 | 0.43±0.018 | 27.7±0.6 | 1.8±0.5 | – | – | – | – | 1455 | | S13W90 |
| 61 | 02 Aug 19 | 1145 | 9.25 | 1.26±0.044 | 0.522±0.024 | 46.5±1.3 | 0.36±0.26 | 549 | 221 | 102 | 1.2 | 1030 | | S12W26 |
| 62 | 02 Aug 19 | 2230 | 10.5 | 1.95±0.07 | 0.681±0.033 | 36.1±1.1 | 1.7±0.7 | 712 | 228 | 66 | 1.9 | 2100 | | S11W32 |
| 63 | 02 Aug 20 | 0945 | 38.3 | 1.79±0.036 | 0.858±0.021 | 25.1±0.44 | 0.66±0.21 | 1099 | 237 | 122 | 1.6 | 0825 | | S11W38 |
| 64 | 02 Sep 29 | 1300 | 19 | 0.98±0.20 | 0.92±0.21 | 49±8 | 0±7.6 | 277 | 251 | 76 | 4.7 | 0750 | | – |
| 65 | 02 Oct 20 | 0000 | 15 | 0.57±0.11 | 0.137±0.049 | 70±9 | 6.9±7.0 | 1076 | 247 | 21 | 3.0 | 2115* | | S13W48 |
| 66 | 02 Dec 12 | 1415 | 9.75 | 0.63±0.15 | 0.56±0.15 | 64±11 | 0±8.5 | 723 | 287 | 51 | 1.9 | 1235 | | N16W36 |
| 67 | 03 Jun 9 | 1615 | 19.8 | 2.02±0.52 | 1.06±0.32 | 100±22 | 0±10 | 749 | 326 | 100 | 5.1 | 1125 | | N12W29 |
| 68 | 03 Jun 10 | 1245 | 23.3 | 0.74±0.09 | 0.44±0.07 | 37.2±3.4 | 0±2.4 | 762 | 328 | 54 | 1.7 | 1105 | | N12W44 |
| 69 | 03 Jun 11 | 2200 | 18 | 0.55±0.09 | 0.57±0.10 | 199±21 | 0±3.5 | NA | | | | 2005 | | N14W57 |
| 70 | 03 Jun 14 | 0430 | 9.5 | 1.65±0.39 | 0.62±0.20 | 59.9±12 | 0±11 | – | – | – | – | 0025 | | – |
| 71 | 03 Jul 11 | 0345 | 6.25 | 1.73±0.51 | 0.41±0.19 | 72±18 | 0±21 | – | – | – | – | – | | – |
| 72 | 03 Jul 11 | 1000 | 8 | 1.67±0.52 | 0.46±0.22 | 74±20 | 0±24 | 388 | 189 | 85 | 2.6 | 0650 | | S10E04 |
| 73 | 03 Aug 19 | 1015 | 43.8 | 0.724±0.054 | 0.264±0.030 | 86.4±4.6 | 0.8±0.8 | 412 | 257 | 35 | 2.7 | 0755 | | S12W63 |
| 74 | 03 Sep 30 | 0130 | 46.5 | 1.24±0.18 | 0.46±0.09 | 22.1±2.6 | 5.3±5.4 | – | – | – | – | 0030 | | N09W40 |
| 75 | 03 Oct 4 | 1515 | 5.25 | 2.47±0.51 | 0.94±0.24 | 20.0±3.7 | 10.6±11 | 1425 | 273 | 17 | 2.2 | 1310 | | – |
| 76 | 03 Oct 4 | 2045 | 25.2 | 0.97±0.07 | 0.436±0.046 | 88.3±5.1 | 3.5±2 | 1262 | 258 | 103 | 2.0 | 1910 | | S06E24 |
| 77 | 03 Oct 22 | 0700 | 25 | 0.926±0.030 | 0.433±0.019 | 52.8±1.3 | 0.58±0.33 | 1163 | 104 | 101 | 3.5 | – | | N07E25 |
| 78 | 03 Oct 25 | 0030 | 13.5 | 1.08±0.049 | 0.423±0.027 | 36.6±1.3 | 0.79±0.56 | – | – | – | – | 2140* | | N05W09 |
| 79 | 03 Dec 30 | 1345 | 16.3 | 0.97±0.19 | 0.70±0.16 | 5.7±0.9 | 0±9.8 | NA | | | | 1150 | | N13W70 |
| 80 | 04 Feb 28 | 0445 | 11.3 | 1.25±0.17 | 0.69±0.11 | 12.8±1.4 | 0±3.7 | 397 | 309 | 85 | 1.9 | 0325 | | N14W47 |
| 81 | 04 Mar 6 | 0715 | 10.8 | 0.49±0.13 | 0.30±0.11 | 28.0±4.9 | 0±15 | 202 | 199 | 35 | 1.3 | 0510 | | – |
| 82 | 04 Mar 30 | 0015 | 35.8 | 0.86±0.09 | 0.45±0.06 | 43.1±3.4 | 0±1.9 | NA | | | | 1955* | | N15E16 |
| 83 | 04 Apr 1 | 0000 | 24 | 0.54±0.09 | 0.233±0.058 | 41.2±4.5 | 0±4.3 | NA | | | | 2005* | | N15W11 |
| 84 | 04 Jun 27 | 0400 | 16 | 1.20±0.16 | 0.55±0.10 | 3.82±0.45 | 0±5.2 | – | – | – | – | 0210 | | – |
| 85 | 04 Jul 22 | 1715 | 3.75 | 1.06±0.14 | 0.646±0.11 | 78±8 | 3.0±3.0 | 574 | 189 | 45 | 6.3 | 1105 | | N03E10 |
| 86 | 04 Jul 22 | 2200 | 8 | 0.68±0.06 | 0.446±0.050 | 81.9±5.2 | 2.6±1.9 | – | – | – | – | – | | – |
| 87 | 04 Jul 23 | 0700 | 6 | 0.499±0.054 | 0.357±0.047 | 106±7 | 0±1.6 | – | – | – | – | – | | – |
| 88 | 04 Jul 23 | 1345 | 23.3 | 0.565±0.021 | 0.397±0.018 | 108±2.7 | 3.3±0.8 | – | – | – | – | – | | – |
| 89 | 04 Jul 24 | 1400 | 26 | 1.05±0.054 | 0.456±0.032 | 82.1±3.3 | 1.7±1.0 | – | – | – | – | – | | – |

| | | | | | | | | | | | | | |
|---|---|---|---|---|---|---|---|---|---|---|---|---|---|
| **90** | 04 Aug 31 | 0730 | 24.5 | 1.14±0.26 | 1.03±0.26 | 28.8±5.2 | 13±13 | 311 | 272 | 70 | 2.7 | 0530 | | N06W82 |
| **91** | 04 Oct 30 | 0545 | 42.3 | 0.637±0.021 | 0.494±0.02 | 81.2±1.9 | 0.58±0.29 | – | – | – | – | 0335 | | N14W15 |
| **92** | 04 Nov  1 | 0800 | 16 | 0.708±0.049 | 0.320±0.031 | 81.6±3.9 | 0.53±0.53 | 925 | 266 | 146 | 2.4 | 0550 | Y | – |
| **93** | 05 Jan 13 | 2100 | 35 | 0.529±0.07 | 0.44±0.07 | 52.6±4.5 | 0±2.2 | 495 | 97 | 360 | 3.8 | <1700 | | S07E16 |
| **94** | 05 May  5 | 2300 | 33 | 0.616±0.018 | 0.339±0.013 | 69.8±1.4 | 0.60±0.27 | 1180 | 82 | 360 | 2.8 | 2015 | | S03W65 |
| **95** | 06 Aug 29 | 2000 | 46 | 0.513±0.045 | 0.328±0.037 | 17.8±1.0 | 2.4±1.7 | NA | | | | 1740 | | S05W39 |
| **96** | 06 Nov 21 | 1015 | 11.8 | 1.34±0.19 | 0.82±0.14 | 3.67±0.47 | 5.0±5.0 | – | – | – | – | 0830 | | S05W37 |
| **97** | 06 Nov 21 | 2100 | 17.0 | 1.27±0.11 | 0.69±0.08 | 10.6±0.7 | 0±1.5 | – | – | – | – | 1825 | | – |
| **98** | 06 Nov 22 | 1500 | 13 | 2.09±0.34 | 1.17±0.22 | 25.3±3.6 | 6.0±6.0 | 1044 | 249 | 74 | 3.6 | 1135 | | – |
| **99** | 10 Feb  8 | 0100 | 23 | 0.541±0.11 | 0.48±0.12 | 39.6±5.6 | 0±6.0 | – | – | – | – | – | | – |
| **100** | 10 Feb 12 | 1730 | 32.5 | 0.69±0.10 | 0.43±0.08 | 68±7 | 0±3.1 | 509 | 45 | 360 | 7.5 | 1125 | | N26E11 |
| **101** | 10 Sep  1 | 0130 | 22.5 | 0.71±0.09 | 0.45±0.07 | 106±9 | 0±2.5 | 1304 | 207 | 360 | 4.8 | 2050* | Y | – |
| **102** | 10 Oct 17 | 1200 | 24 | 1.12±0.24 | 0.78±0.19 | 41±7 | 0±12 | 304 | 269 | 54 | 3.4 | 0855 | | S19W31 |
| **103** | 11 Feb 18 | 0415 | 17.8 | 1.34±0.20 | 0.69±0.13 | 19.1±2.3 | 6.0±5.9 | 490 | 286 | 82 | 6.1 | 2135* | | S20W45 |
| **104** | 11 Sep  7 | 0200 | 33 | 0.66±0.09 | 0.191±0.043 | 143±13 | 0±2.1 | 575 | 302 | 360 | 4.0 | 2220* | Y | N14W18 |
| **105** | 11 Dec 14 | 0700 | 33 | 0.60±0.11 | 0.53±0.11 | 43.6±5.6 | 0±5.0 | – | – | – | – | 0515 | | S18W90 |
| **106** | 12 May 14 | 1130 | 30 | 0.70±0.08 | 0.321±0.053 | 43.8±3.6 | 0±2.1 | 551 | 229 | 48 | 1.8 | 0940 | | N07W45 |
| **107** | 12 Jun  8 | 0915 | 18 | 0.66±0.11 | 0.48±0.10 | 32.4±3.8 | 0±5.1 | 308 | 229 | 34 | 2.3 | 0715 | | N13W40 |
| **108** | 12 Sep 12 | 1100 | 38 | 0.426±0.059 | 0.325±0.055 | 29.6±2.5 | 0±2.3 | 261 | 283 | 66 | 4.4 | 0720 | | – |
| **109** | 13 Jan 21 | 1430 | 17 | 1.39±0.29 | 0.44±0.14 | 6.87±1.2 | 0±9.5 | – | – | – | – | 1150 | | – |
| **110** | 13 Apr 22 | 0600 | 15 | 0.829±0.19 | 0.38±0.12 | 136±23 | 0±6.8 | – | – | – | – | – | | – |
| **111** | 13 Apr 24 | 0600 | 24 | 1.01±0.17 | 0.60±0.13 | 35.2±4.6 | 0±5.8 | – | – | – | – | – | | – |

\* DH III burst and flare occurred on the day prior to the date shown for the SEP event onset.